\documentclass[a4paper]{aa}
\usepackage{graphicx}

\def \rsun {\ifmmode$R$_{\odot}\else R$_{\odot}$\fi}
\def \nh {N${\rm _H}$}
\def \hcm {\hbox {\ifmmode $ atom cm$^{-2}\else atom cm$^{-2}$\fi}}
\def \src {X\,1822-371}
\def\approxgt{\mathrel{\hbox{\rlap{\lower.55ex \hbox {$\sim$}}
        \kern-.3em \raise.4ex \hbox{$>$}}}}
\def\approxlt{\mathrel{\hbox{\rlap{\lower.55ex \hbox {$\sim$}}
        \kern-.3em \raise.4ex \hbox{$<$}}}}
\newcommand{\mc}{\multicolumn}

\newcommand {\chisq} {$\chi ^{2}$}

\begin{document}

\title{Broad-band BeppoSAX observation of the low-mass X-ray
binary X\,1822-371}

\author{A.N. Parmar\inst{1} \and T. Oosterbroek\inst{1} 
\and S. Del Sordo\inst{2} \and A. Segreto\inst{2} \and A. Santangelo\inst{2}
\and D. Dal Fiume\inst{3} \and M.~Orlandini\inst{3}}

\offprints{A.N. Parmar (aparmar@astro.estec. esa.nl)}

\institute{Astrophysics Division, Space Science Department of ESA, 
ESTEC, P.O. Box 299, 2200 AG Noordwijk, The Netherlands
\and IFCAI, CNR, via La Malfa 153, I-90146 Palermo, Italy
\and Istituto TESRE, CNR, via Gobetti 101, I-40129 Bologna, Italy}

\thesaurus{(02.01.2; 08.09.2; 08.14.1; 13.25.5)}

\date{Received: 1999 November 10; Accepted: 2000 January 13}

\authorrunning{A.N. Parmar et al.}

\maketitle 

\markboth{BeppoSAX observation of \src}
{BeppoSAX observation of \src}

\begin{abstract}

Results of a 1997 September 9--10 BeppoSAX observation 
of the 5.57~hr low-mass X-ray binary (LMXRB)
\src\ are presented.
The 0.3--40~keV spectrum is unusually complex and cannot be fit by any of
the standard models applied to other LMXRB.
At least two components are required. One component has a shape
consistent with that expected from the Comptonization of an input
soft (Wien) spectrum while the other, contributing $\approxgt$40\%
of the 1--10~keV flux, is consistent with being a blackbody.
In addition, there
is a ``dip'' in the spectrum
which can be modeled by a $1.33 \pm ^{0.05} _{0.11}$~keV absorption edge with
an optical depth, $\tau$, of $0.28 \pm 0.06$. 
If the same model is fit to ASCA Solid-State Imaging
Spectrometer spectra obtained
in 1993 and 1996, then reasonable fits are also obtained,
with a similar absorption feature required. 
The nature of this feature is highly uncertain; its energy corresponds
to the K-edges of highly ionized Ne~{\sc x} and neutral Mg, or to an L-edge
of moderately ionized Fe. Surprisingly, no strong ($\tau > 0.05$) Fe-K or
($\tau > 0.18$) O-K edges are visible. The folded lightcurve of \src\ is
similar to previous observations, except that no strong softening
is seen near the eclipse. An updated orbital ephemeris is provided.


\end{abstract}

\section{Introduction}
\label{sect:introduction}

The 5.57~hr low-mass X-ray binary (LMXRB) \src\ is viewed almost edge
on with the central X-ray source hidden by material in the orbital plane.
The X-ray lightcurve shows a partial eclipse and a smooth broad modulation
with a minimum just prior to the eclipse (White et al. \cite{w:81}).
The partial nature of the eclipse indicates that the X-ray emitting region is 
extended, and that the observed X-rays are scattered in an 
accretion disk corona, or ADC (White \& Holt \cite{w:82}). 
The optical lightcurve has a similar morphology except that the
eclipse is broader. White \& Holt (\cite{w:82}) showed that the broad X-ray 
modulation can be modeled as obscuration of the ADC by 
the rim of the accretion disk whose thickness is greatest near phase,
$\Phi$, 0.8 (where $\Phi = 0.0$ is mid-eclipse), and least near $\Phi = 0.2$.
Modeling of multi-wavelength lightcurves reveals
that the effective diameter of the ADC is $3 \times 10^{10}$~cm,
about half that of the optically emitting disk, and that structure
in the disk can reach a height of $1.5 \times 10^{10}$~cm
(White \& Holt \cite{w:82}; Mason \& C\'ordova \cite{mc:82};
Hellier \& Mason \cite{h:89}; Puchnarewicz et al. \cite{p:95}).  
No pulsations or bursts have been detected from \src\ (e.g., 
Hellier et al. \cite{h:92}). The 
probable distance to \src\ is 2--3~kpc and the isotropic luminosity
$10^{36}$(d$_2$)$^2$~erg~s$^{-1}$ (Mason \& C\'ordova \cite{mc:82}), where
d$_2$ is distance in units of 2~kpc.
The mean \src\ ${\rm L_x/L_{opt}}$ ratio of 20 compared
to the average for LMXRB of 500 (van Paradijs \&
McClintock \cite{v:95}) implies a unobscured
luminosity of $2 \times 10^{37}$(d$_2$)$^2$~erg~s$^{-1}$, similar
to that observed from the brighter X-ray burst sources.

The X-ray spectrum of \src\ is complex with different results being
obtained from different measurements. 
The 2--40~keV HEAO-1 A2 spectrum can be fit by 
a power-law with a photon index, $\alpha$, of 1.3, a 
high-energy cutoff at 18~keV, and a broad Fe-K line with a 
full-width half maximum (FWHM) of 4~keV. Below 2~keV there is
evidence for an excess in the {\it Einstein} Solid State Spectrometer
spectrum which White et al. (\cite{w:81}) can
model as a 0.25~keV thermal bremsstrahlung, a 0.16~keV blackbody or a
350~eV equivalent width (EW) emission feature at 0.8~keV. 
The combined 0.04--20~keV 
EXOSAT Channel Multiplier Array and Medium Energy Detector Array
spectrum is fit by a power-law with $\alpha = 0.8$ together with a blackbody
with kT = 1.8~keV and a $\sim$1~keV FWHM Fe-K line
(Hellier \& Mason \cite{h:89}). Subsequently, Hellier et al. (\cite{h:92})
obtained a {\it Ginga} 1.5--30~keV spectrum, but were unable to find
a satisfactory fit, although the model used for the EXOSAT spectrum gives
the best result. 

White et al. (\cite{w:97}) report on an ASCA observation
of \src. The best-fit Solid-State Imaging Spectrometer (SIS) and
Gas Imaging Spectrometer (GIS) 0.5--10~keV spectrum is also complex and
cannot be easily fit by any of the standard models. White et al. (\cite{w:97})
characterize the spectrum using a power-law continuum 
with $\alpha \sim 0.52$. Structured residuals remain with a pronounced
dip at $\sim$1.5~keV and a strong decrease $\approxgt$7~keV. There are
no emission features present other than two weak features at 6.4~keV
and 7.1~keV, consistent with Fe K$\alpha$ and K$\beta$ lines. The EW
of the K$\alpha$ line is $\sim$40~eV. However, as
White et al. (\cite{w:97}) note, the observed
K$\beta$/K$\alpha$ ratio of $\sim$40\% does not match the theoretical
value of 13\%. White et al. (\cite{w:97}) attempt to model the reduction
in observed flux $\approxgt$7~keV by an Fe-K absorption edge, but find
that edges from a range of ionization states are necessary to explain the
observed shape.

As White et al. (\cite{w:97}) point out, accurately measuring the continuum
$\approxgt$10~keV is vital in order to constrain the Fe-K absorption
feature seen in the ASCA spectrum. Indeed, these authors used 
non-simultaneous EXOSAT and {\it Ginga} data to try to do this, but
were unsuccessful. We report here on a BeppoSAX observation of \src\ 
where the 0.3--40~keV
spectrum was observed simultaneously with good sensitivity.
We compare the results of the BeppoSAX spectral fits with those
obtained by White et al. (\cite{w:97}) and those obtained from a 
subsequent ASCA observation.

\begin{figure*}
\hbox{\hspace{0.8cm}
\includegraphics[height=16.0cm,angle=-90]{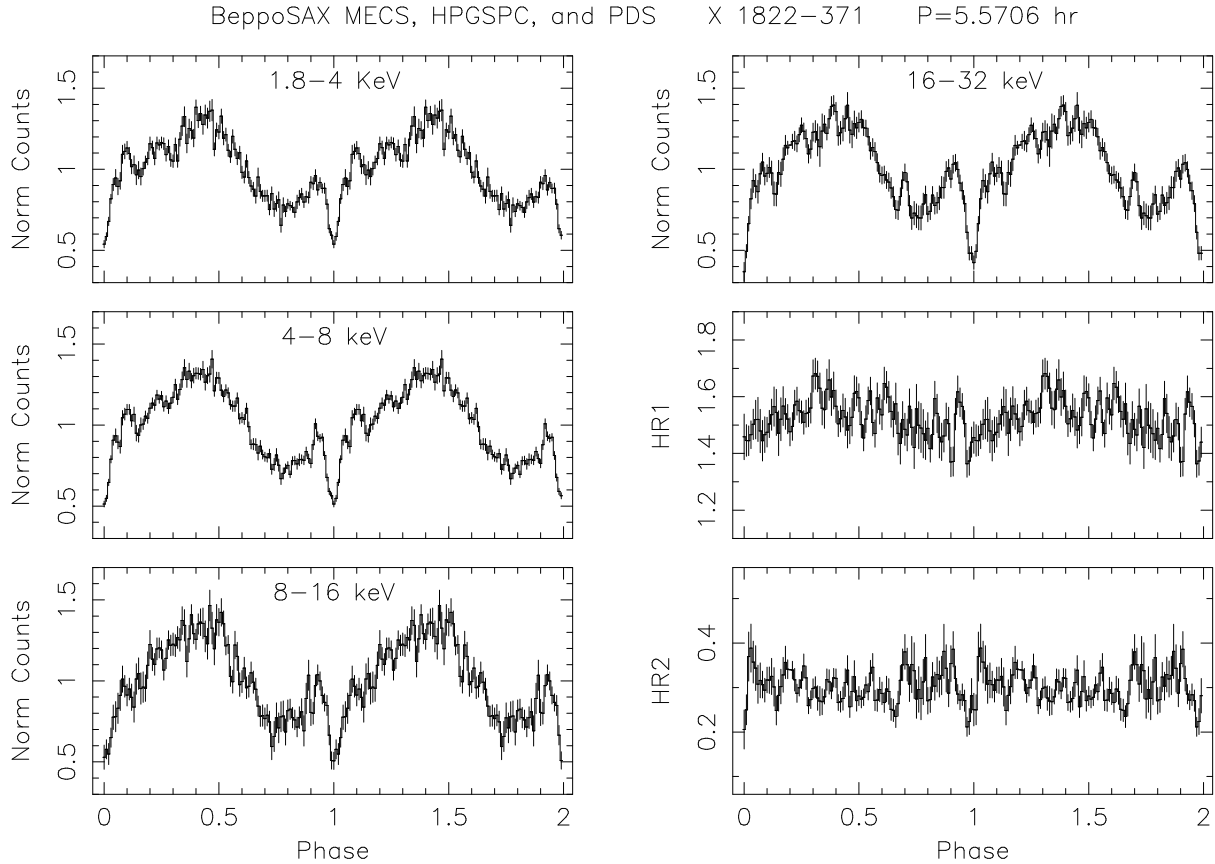}}
  \caption[]{The \src\ folded lightcurves and hardness ratios. The 
             energy ranges are indicated. The upper two left panels
             have 100 phase bins, the others 50. HR1 and HR2 are the 
             4--8~keV/1.8--4~keV and 16--32~keV/10--16~keV
             hardness ratios, respectively. The ordinate extrema of
             each lightcurve are normalized to
             the mean count rate for ease of comparison}
  \label{fig:folded}
\end{figure*}

\section{Observations}
\label{sect:observations}

Results from the Low-Energy Concentrator Spectrometer (LECS;
0.1--10~keV; Parmar et al. \cite{p:97}), the Medium-Energy Concentrator
Spectrometer (MECS; 1.8--10~keV; Boella et al. \cite{b:97}),
the High Pressure Gas Scintillation Proportional Counter
(HPGSPC; 5--120~keV; Manzo et al. \cite{m:97}) and the Phoswich
Detection System (PDS; 15--300~keV; Frontera et al. \cite{f:97}) on-board
BeppoSAX
are presented. All these instruments are coaligned and collectively referred
to as the Narrow Field Instruments, or NFI.
The MECS consists of two grazing incidence
telescopes with imaging gas scintillation proportional counters in
their focal planes. The LECS uses an identical concentrator system as
the MECS, but utilizes an ultra-thin entrance window and
a driftless configuration to extend the low-energy response to
0.1~keV. The non-imaging HPGSPC consists of a single unit with a collimator
that was alternatively rocked on- and 180\arcmin\ off-source 
every 96~s during the observation. 
The non-imaging
PDS consists of four independent units arranged in pairs each having a
separate collimator. Each collimator was alternatively
rocked on- and 210\arcmin\ off-source every 96~s during 
the observation.

The region of sky containing \src\ was observed by BeppoSAX
on 1997 September 09 12:24 UT to September 10 11:47 UT.
Good data were selected in the standard way
using the SAXDAS 2.0.0 data analysis package.
LECS and MECS data were extracted centered on the (on-axis) 
position of \src\ 
using radii of 8\arcmin\ and 4\arcmin, respectively.
The exposures 
in the LECS, MECS, HPGSPC, and PDS instruments are 14.8~ks, 37.0~ks,
19.5~ks, and 18.7~ks, respectively. 
Background subtraction for the imaging instruments
was performed using standard files, but is not critical for such a
bright source. 
Background subtraction for the HPGSPC was 
carried out using data obtained when the instrument
was looking at the dark Earth and for the PDS using data
obtained during intervals when the collimator was offset from the 
source. 

The BeppoSAX data is compared with results from the Solid State Imaging
Spectrometers SIS0 and SIS1 (0.6--10~keV), 
on-board ASCA (Tanaka et al. \cite{t:94}). The energy resolution
of the SIS is a factor of a few better, except at
the lowest energies, than that of the LECS and MECS. 
ASCA observed \src\ twice. The
first observation took place between 1993 October 07 03:33 and 23:55~UTC
and the second between 1996 September 26 05:44 and September 27 08:05~UTC.
The SIS exposures for each observation are 36.6~ks and 26.0~ks both
using 1-CCD BRIGHT mode. All data were screened and processed using the
standard Rev2 pipeline. The source count rates of 
$\approxlt$6.5~s$^{-1}$~SIS$^{-1}$ means that pulse pile-up is unlikely
to be significant.

\section{Analysis and results}
\label{sect:analysis}

\subsection{BeppoSAX lightcurve}
\label{subsect:lc}

Background subtracted lightcurves in the energy ranges 1.8--4~keV (MECS),
4--10~keV (MECS), 10--16~keV (HPGSPC) and 16--32~keV (PDS) were
extracted and folded using the linear
ephemeris of Hellier \& Smale (\cite{h:94}).
Fig.~\ref{fig:folded} shows these lightcurves together with 
two hardness ratios (4--10~keV counts divided by 1.8--4~keV counts
and 16--32~keV counts divided by 10--16~keV counts).
The lightcurves are similar to those reported previously,
showing gradual reductions 
in flux which reach minima around $\Phi \sim 0.8$,
before the partial eclipse and rapid increases following
egress. Just prior to the partial eclipse each lightcurve
shows a narrow increases in flux. 
The depth of the partial eclipse increases
with increasing energy.
There are no strong variations in hardness ratio. 
However, the
4--8~keV/1.8--4~keV hardness ratio (HR1) shows a broad 
variation with a maximum around $\Phi \sim 0.4$, as well as
a narrow increase just prior to eclipse.
Instead, the 16--32~keV/8--16~keV hardness ratio (HR2) does not
vary appreciably with phase, except 
between $\Phi = 0.7$--1.1 where a harder interval, punctuated
by the eclipse, is present. HR1 is similar in shape to the
6--30~keV/1--6~keV hardness ratio obtained from {\it Ginga} observations
(see Fig.~1 of Hellier et al. \cite{h:92}), except that no strong softening
is visible in the BeppoSAX data during the eclipse. This is in
contrast to previous results, where the eclipse depths measured by
EXOSAT and {\it Ginga} both showed a similar strong dependence on
energy (Hellier et al. \cite{h:92}).
The lack of any large change in the BeppoSAX hardness ratios 
justifies the use of the entire data set in
the spectral analysis in Sect.~\ref{subsect:spec}.

\subsection{Eclipse timing}

Eclipse arrival times using 
recent BeppoSAX MECS, ASCA and RXTE 
Proportional Counter Array data have been determined. 
Since there are substantial gaps in 
the data and the count rates low, 
the BeppoSAX and ASCA lightcurves were first folded
on the 5.5706~hr period
of Hellier \& Smale (\cite{h:94}). A model consisting
of a Gaussian and a constant was then fit
to the eclipse phases ($\Phi = 0.95$--1.05) of the folded
lightcurves. 
The arrival time of the eclipse was then assigned to 
the eclipse which occurred closest to mid-time of the observation. For the
MECS the uncertainty on the arrival time was obtained directly from
the fit. Data from the 4 ASCA instruments were analyzed independently
and the uncertainties in the arrival times were derived from the 
spreads in the
obtained values, while the arrival times were defined as the averages
from the 4 instruments. 
The RXTE observation was made between 1996
September 26 16:03 and September 27 07:19~UTC. Since only
one eclipse was observed, the Gaussian and a constant model 
was fit directly to the eclipse interval in the lightcurve.
The uncertainty in the RXTE arrival
time was estimated by comparing results obtained
using (1) a range of eclipse phases around 
those given above using the Gaussian and constant model  
and (2) including a linear term in the above model
and fitting over the same range of phases. 
The spread in values obtained using these two methods
was larger than the uncertainties
obtained from the individual fits, and so was adopted as an estimate of the 
overall uncertainty.
The fit results, which extend
the measurements in Hellier \& Smale (\cite{h:94}) by some 5~yrs,
or 8000 cycles, are summarized in Table \ref{tab:eclipse_times}.

\begin{figure}
\centerline{
\includegraphics[width=8.5cm]{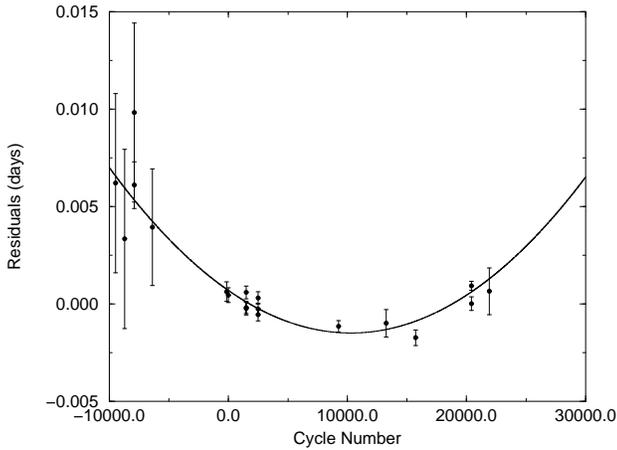}}
\caption[]{The \src\ partial eclipse timing residuals with respect to 
the best-fit linear ephemeris 
(small filled circles). The quadratic ephemeris (solid line) is also
indicated}
\label{fig:residuals}
\end{figure}

The newly determined arrival times together with the
measurements tabulated in Hellier \& Smale (\cite{h:94}) were
fit to obtain an updated ephemeris. Both linear
(\chisq\ = 98.5 for 17 dof) and quadratic ephemerides (\chisq\ 
= 21.4 for 16 dof) were used. The difference in reduced $\chi ^2$
indicates that the
quadratic term in the ephemeris is established with 99.9999\%
confidence. Fig.~\ref{fig:residuals} shows the residuals with
respect to both ephemerides.
The updated quadratic ephemeris is given by:

\begin{eqnarray}
{\rm T_{ecl}} & = & 2445615.30964(15) + 0.232108785(50)\,{\rm N} \nonumber  \\
             &   & + {\rm 2.06(23)\times 10^{-11} \,N^{2}.} \nonumber
\end{eqnarray}

\begin{table}
\begin{center}
\caption[]{New X-ray eclipse times for \src. A compilation of previous
measurements can be found in Hellier \& Smale (\cite{h:94}). 
Uncertainties are given at 68\% confidence}
\begin{tabular}{llll}
\hline
\noalign {\smallskip}
JD$_{\odot}$ & Uncertainty & Cycle & Satellite \\
\hline
\noalign {\smallskip}
2,449,268.00984  & 0.00040 & 15737 & ASCA \\
2,450,353.35425  & 0.00035 & 20412 & ASCA \\
2,450,353.58728  & 0.00023 & 20414 & RXTE \\
2,450,701.51870  & 0.00120 & 21912 & BeppoSAX \\
\noalign {\smallskip}                       
\hline
\label{tab:eclipse_times}
\end{tabular}
\end{center}
\end{table}

\subsection{BeppoSAX spectrum}
\label{subsect:spec}

The overall spectrum of \src\ was first investigated by simultaneously
fitting data from all the BeppoSAX NFI. 
All spectra were rebinned using standard procedures.
Data were selected in the energy ranges
0.3--4.0~keV (LECS), 2.0--10~keV (MECS), 7.0--30~keV (HPGSPC),
and 15--40~keV (PDS) 
where the instrument responses are well determined and sufficient
counts obtained. 
This gives
background-subtracted count rates of 1.0, 4.2, 8.8 and 4.0~s$^{-1}$ 
for the LECS, MECS, HPGSPC, and PDS, respectively.
The photo-electric absorption
cross sections of Morisson \& McCammon (\cite{m:83}) and the
abundances of Anders \& Grevesse (\cite{a:89}) are used throughout.
Factors were included in the spectral fitting to allow for normalization 
uncertainties between the instruments. These factors were constrained
to be within their usual ranges during the fitting. All spectral uncertainties
and upper-limits are given at 90\% confidence.

\begin{table}
\begin{center}
\caption[]{Best-fit to the BeppoSAX NFI spectrum. The model consists 
of absorbed blackbody and {\sc comptt} continuum, a Gaussian emission 
line, and an absorption edge}
\begin{tabular}{ll}
\hline
\noalign {\smallskip}
Parameter              &  \\
\hline
\noalign {\smallskip}
\nh\ (atom cm$^{-2}$)    & $(1.2  \pm _{1.2} ^{0.8}) \times 10^{20}$ \\ 
Blackbody kT (keV)       & $1.88 \pm 0.02$ \\
kT${\rm _e}$ (keV)        & $4.52 \pm 0.02$ \\
${\rm \tau_p}$           & $26.1 \pm 0.5$ \\
kT${\rm _W}$  (keV)       & $0.15 \pm 0.02$ \\
Line energy (keV)        & $6.57 \pm ^{0.04} _{0.07}$ \\ 
Line FWHM (keV)          & $0.69 \pm 0.19 $ \\
Line EW (keV)            & $0.150 \pm 0.025$ \\
Edge energy (keV)        & $1.33 \pm ^{0.05} _{0.11}$ \\ 
Edge $\tau$              & $0.28 \pm 0.06 $ \\
$\chi ^2$/dof            & 122.1/125 \\
\noalign {\smallskip}                       
\hline
\label{tab:spec_paras}
\end{tabular}
\end{center}
\end{table}

\begin{figure}
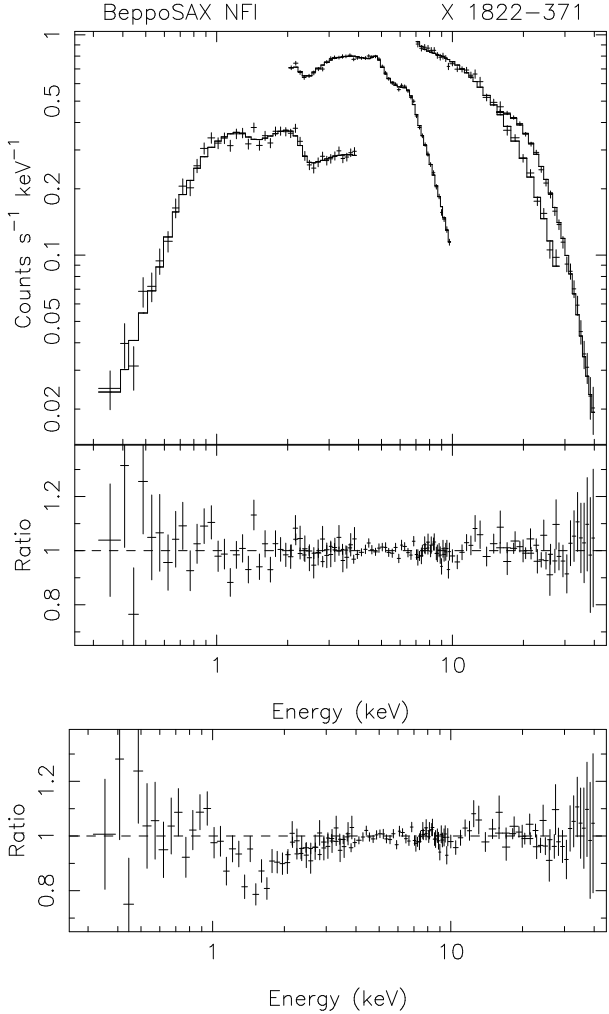

\hbox{\hspace{0.0cm}\includegraphics[height=8.0cm,angle=-90]{h1846f3a.ps}}
\vspace{-0.0cm}
\hbox{\hspace{0.0cm}\includegraphics[height=8.0cm,angle=-90]{h1846f3b.ps}}
  \caption[]{The \src\ count spectrum and residuals when a {\sc comptt} and 
             blackbody continuum with an edge and an 
             Fe line are fit (Table~\ref{tab:spec_paras}). 
             The middle panel shows the ratio of the 
             data divided the model and the lower panel shows the same 
             when the edge is excluded}
  \label{fig:spectrum}
\end{figure}

\begin{figure}
  \centerline{
   \includegraphics[width=9.0cm,angle=-90]{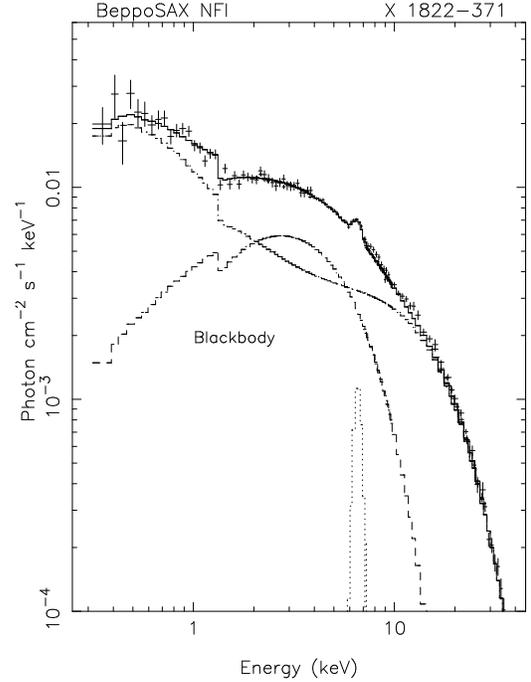}}
  \caption[]{The overall \src\ NFI photon spectrum derived using the 
             {\sc comptt} and blackbody continuum with 
             an edge and an Fe line. The individual contributions
             are indicated}
  \label{fig:photons}
\end{figure}

Initially, a standard LMXRB spectral model consisting of a
a cutoff power-law (${\rm E^{-\alpha}\exp-(E_{c}/kT)}$) appropriate
to low-luminosity sources was fit to the BeppoSAX spectrum.
Since the central X-ray source in \src\ is hidden from direct view 
the overall luminosity of \src\ is highly uncertain, and an
additional blackbody component, frequently observed in more luminous 
LMXRB (e.g., White et al. \cite{w:88}) was also included.
The cutoff power-law model gives an unacceptable fit with a $\chi ^2$ of
670 for 133~dof. Adding a 1.5~keV blackbody reduces the $\chi ^2$ to
596 for 131~dof. Inspection of the residuals indicated the presence
of an Fe-K line. Adding a broad line at $\sim$6.5~keV 
further reduces the $\chi^2$ to 408 for 128~dof. 
The remaining residuals are indicative of an intense low-energy 
excess. We attempted to model this feature using a Gaussian emission
line, thermal bremsstrahlung, blackbody and power-law 
components. Only the fit including a Gaussian feature was
acceptable with a $\chi^2$ of 144.5 for 125~dof. 
Thus, this
model consists of a cutoff power-law with $\alpha = -0.78 \pm 0.02$
and ${\rm E_c} = 5.69 \pm 0.06$~keV 
and a $1.27 \pm 0.03$~keV blackbody continuum together
with two Gaussian emission features.
The nature of the low-energy feature is highly uncertain.
Its energy can only be constrained to be $<$0.50~keV,
it is extremely broad (FWHM = 1.5~keV), and its EW is implausibly 
high ($\sim$7~keV).
We therefore reject this model as being physically unreasonable. 

The HEAO-1 \src\ spectrum of White et al. (\cite{w:81}) is well fit
above 2~keV using a power-law model 
with a high-energy break (${\rm \exp((E_{cut}-E)/E_{fold})}$ for 
${\rm E > E_{cut}}$) at 17~keV. Fitting 
this model to the BeppoSAX NFI spectrum does not produce
an acceptable fit with a $\chi ^2$ of 1960 for 132~dof. If a 
blackbody is included, then the situation improves with a $\chi ^2$ of 
342 for 130~dof. Including a broad Fe line decreases the $\chi ^2$ to
198.9 for 127~dof. The major contribution to $\chi ^2$ comes from
the region near 1~keV where the model overestimates the observed
spectrum. We attempted to model feature this using partial covering
(the {\sc pcfabs} model in {\sc xspec}, an ionized absorber 
the {\sc absori} model in {\sc xspec}),
and an absorption edge. Only the edge was successful with an
energy of $1.30 \pm 0.07$~keV and an optical depth, $\tau$, of
$0.35 \pm 0.09$ for a $\chi ^2$ of 139.4 for 125~dof.
The best-fit value of ${\rm E_{cut}}$ of $17.1 \pm 0.7$~keV is in
good agreement with that obtained from HEAO-1 of $17.4 \pm 0.5$~keV
(White et al. \cite{w:81}).
There is no evidence for the presence of low-energy thermal
components, Fe-L emission near 1~keV as proposed
by White et al. (\cite{w:81}), or a reflection component 
(the {\sc pexrav} model in {\sc xspec}).

Finally, we investigated whether Comptonization models such as
the {\sc xspec} model {\sc comptt} (Titarchuk \cite{ti:94};
Hua \& Titarchuk \cite{h:95}; Titarchuk \& Lyubarskij \cite{t:95}),
which self-consistently calculate the spectrum produced
by the Comptonization of soft photons in a hot plasma, could
be applicable to \src. This model
contains as free parameters the temperature of the Comptonizing
electrons kT${\rm _e}$, the plasma optical depth with respect to
electron scattering
${\rm \tau_p}$ and the input 
temperature of the soft photon (Wien) distribution kT${\rm _W}$. A
spherical geometry was assumed for the comptonizing region.
This replaced the power-law and high-energy cutoff in the previous model
and the fit repeated. A good fit is obtained with a $\chi ^2$
of 122.1 for 125 dof.
The 1--10~keV flux is $5.0 \times 10^{-10}$~erg~cm$^{-2}$~s$^{-1}$
and the blackbody contributes 43\% of the total flux in this energy range.  
The best-fit parameters are given in Table~\ref{tab:spec_paras}.
If a disk geometry is assumed for the Comptonizing region, 
the best-fit parameters
are essentially unchanged except for $\tau _p$ which reduces to $13.3 \pm 0.3$.
Fig.~\ref{fig:spectrum} shows the count spectra and the data/model
ratios with, and without, the edge. The depth of the feature ($\sim$20\%) is
significantly greater than the uncertainty in LECS calibration at
the corresponding energies ($\approxlt$5\%).
Fig.~\ref{fig:photons} shows the
incident photon spectrum and illustrates how fitting
a power-law in the energy range 1.0--10~keV would produce a  
``dip'' at $\sim$1.5~keV as observed by White et al. (\cite{w:97}).
The 90\% confidence upper limits to the optical
depths of any O-K and Fe-K edges at 0.54~keV and 7.1~keV are 0.18 and
0.05, respectively. 

\subsection{ASCA spectrum}
\label{subsect:asca}

We next examined whether the BeppoSAX best-fit model presented above
is consistent with results from ASCA.
For the 1993 ASCA observation these are presented in 
White et al. (\cite{w:97}). A simple power-law was first fit to the ASCA
SIS spectra confirming that the overall shape of the spectrum is as 
reported in
White et al. (\cite{w:97}) when using the latest processing.
Next, the best-fit BeppoSAX model 
was fit to the SIS spectra.
For the fit to the 1993 spectra an additional line feature
at $\sim$7~keV was included.  
The results are presented in Table~\ref{tab:asca}
and show that the best-fit BeppoSAX model also provides a reasonable fit
to the ASCA SIS spectra. 
During the 1993 and 1996 observations,
the 1--10~keV fluxes were 
$5.8 \times 10^{-10}$~erg~cm$^{-2}$~s$^{-1}$
and $5.3 \times 10^{-10}$~erg~cm$^{-2}$~s$^{-1}$, 
and the blackbody contributed 48\% and 47\% of the flux
in this energy range, respectively.

Comparing the best-fit parameters with those obtained with BeppoSAX
reveals the following: (1) The overall spectral shape 
is similar in all three observations and may be modeled using 
{\sc comptt} and blackbody continuum components. 
(2) There is a substantial (EW 
$\sim$100--300~eV) Fe-K
feature present which may be modeled as a broad feature at $\sim$6.5~keV,
or as the sum of two narrower features at $\sim$6.4~keV and $\sim$7.0~keV.
(3) Using the relation between
optical and X-ray extinctions of Predehl \& Schmitt (\cite{ps:95}),
all three best-fit absorptions 
are inconsistent with the value derived from the color excess of
0.13 in Mason et al. (\cite{m:82}) of $8\times 10^{20}$~atom~cm$^{-2}$.
(4) All three spectra reveal clear evidence for a ``dip'' at $\sim$1.3~keV,
which may be modeled by an absorption edge. The energies of this
feature ($1.33 \pm ^{0.05} _{0.11}$~keV, 
$1.37 \pm 0.03$~keV, and $1.28 \pm 0.03$~keV)
are in reasonable agreement, while there are clear variations
in its optical depth ($\tau = 0.28 \pm 0.06$, 
$0.093 \pm 0.012$, and $0.15 \pm 0.02$).
 
\begin{table}
\begin{center}
\caption[]{The best-fit BeppoSAX spectral model (see Sect.~\ref{subsect:spec}) 
applied to the ASCA SIS spectra}
\begin{tabular}{lll}
\hline
\noalign {\smallskip}
Parameter              &  \mc{2}{c}{Observation Date \hfil}\\
                       & \hfil 1993 & \hfil 1996 \\
\hline
\noalign {\smallskip}
\nh\ (atom cm$^{-2}$)    & $<$$3 \times10^{19}$ 
                         & $(5.0 \pm 0.5) \times 10^{20}$ \\ 
Blackbody kT (keV)       & $1.79 \pm 0.02$  & $1.66 \pm 0.02$ \\
kT${\rm _e}$ (keV)        & $7.2 \pm 0.6$ & $9.9 \pm 0.9$\\
${\rm \tau_p}$           & $23.3 \pm 0.15$ & $22.8 \pm 0.3$\\
kT${\rm _W}$  (keV)       & $0.153 \pm 0.010$ & $0.174 \pm 0.013$\\
Line energy (keV)        & $6.46 \pm 0.03$ & $6.49 \pm 0.08$\\ 
Line FWHM (keV)          & $0.51 \pm 0.08$ & $1.65 \pm 0.25$ \\
Line EW (keV)            & $0.14 \pm 0.02$ & $ 0.285 \pm 0.040$\\
Line energy (keV)        & $7.06 \pm ^{0.02} _{0.04}$ & \dots\\ 
Line FWHM (keV)          & $<$$0.30$  & \dots \\
Line EW (keV)            & $0.065 \pm 0.010$ & \dots \\
Edge energy (keV)        & $1.37 \pm 0.03$ & $1.28 \pm 0.03$ \\ 
Edge $\tau$              & $0.093 \pm 0.012$ & $0.15 \pm 0.02$ \\
$\chi ^2$/dof            & 766.0/570 & 721.5/559\\
\noalign {\smallskip}                       
\hline
\label{tab:asca}
\end{tabular}
\end{center}
\end{table}

\section{Discussion}
\label{sect:discussion}

We present results of a 1997 September BeppoSAX observation of \src\
and compare these with earlier ASCA observations when the source
had a similar 1.0--10~keV intensity. 
The spectrum is unusually complex and cannot be 
fit by any of the usual models applied to LMXRB 
such as a cutoff power-law and blackbody unless an unusually strong
low-energy emission feature is included.
A good fit is obtained to the 0.3--40~keV 
BeppoSAX spectrum with the combination of a Comptonization
component and a blackbody together with an Fe-K emission line 
and an absorption edge.
The same model provides reasonable fits to ASCA SIS spectra with a
similar absorption feature being required.

There are at least three highly unusual features of the 
\src\ spectrum. The first, also pointed out by White et al. (\cite{w:97}),
is that the continuum is much harder than is typical of similar
luminosity LMXRBs, where $\alpha$ is usually
1.5--2.5. The spectrum scattered in an ADC is expected
to resemble the original, unless the optical depth is large.
The second is the extremely large contribution of
the blackbody component ($\approxgt$40\% of the total), whereas fits 
to luminous LMXRB indicate the presence of blackbodies with 
typical luminosities of 16--34\% of the non-thermal component 
(White et al. \cite{w:88}). However, we note that another LMXRB, the X-ray dip
source XB~1746-371 located in the globular cluster NGC\thinspace6441,
also recently been found to have a strong blackbody-like component which
contributes 88\% of the 1--10~keV flux (Parmar et al. \cite{p:99};
see also Guainazzi et al. \cite{g:99}). 
The blackbody component most probably originates in an optically thick
boundary layer between the accretion disk and the neutron star surface,
or from the neutron star itself. It is therefore unlikely to be
observed directly in \src\ and so it is surprising that it appears 
so bright in comparison with the non-thermal component. The third
unusual aspect of the spectrum is the presence of the strong
low-energy feature, which can be 
modeled as an absorption edge. The fact that a similar component is also
seen in the two ASCA observations implies that this is a stable feature
of the spectrum, but not necessarily that it is correctly modeled.
However, its nature is highly uncertain. The energy corresponds
to K-edges of highly ionized Ne~{\sc x} and neutral Mg, or to an L-edge
of moderately ionized Fe. Surprisingly, no strong ($\tau > 0.05$) Fe-K or
($\tau > 0.18$) O-K edges are evident in the spectrum.

Models for the spectrum of \src\ which involve a significant Comptonized 
component appear plausible. 
A plasma with kT${\rm _e} \sim$5--10~keV and
a ${\rm \tau_p}$ of $\sim$22--26 is
required. These values imply a Comptonization parameter 
${\rm y = 4kT_e \tau_p^2/m_e c^2}$ of $\sim$25--40. Values of y$>$12 imply
that the emerging spectrum will be saturated and have a
Wien-like shape (e.g., Titarchuk \cite{t:94}). 
Guainazzi et al. (\cite{g:99}) show that when the BeppoSAX spectra of
a number of LMXRB located in globular clusters are fit with the
same continuum model as applied here (a blackbody and {\sc comptt}),
then the derived values of ${\rm \tau_p}$ and ${\rm kT_e}$ are respectively 
correlated and anti-correlated with the source luminosity.
Guainazzi et al. (\cite{g:99}) suggest that these correlations may
be qualitatively explained if the X-ray emission at the boundary
layer between the accretion disk and the neutron star surface
is proportional to the accretion rate. If this results in 
an increase in the Comptonizing plasma optical depth this would allow
Compton cooling to become more efficient, 
yielding a lower Comptonizing electron temperature. 
If the intrinsic luminosity
of \src\ is assumed to be $2 \times 10^{37}$~erg~s$^{-1}$ (see Sect.~1), 
then the values derived here using BeppoSAX for ${\rm \tau_p}$ and ${\rm kT_e}$
are in good agreement with the relations derived from the study of the
globular cluster LMXRB X-ray sources. 

The overall X-ray spectrum of \src\ remains poorly understood. In particular,
calculations by e.g., Ko \& Kallman (\cite{k:94}) and Kallman et al.
(\cite{k:96}) show that photo-ionized ADC should be a rich source 
of line emission, which does not appear to be the case here as only
a moderate EW Fe-K line is seen. 
Vrtilek et al. (\cite{v:93}) discuss the situation where an X-ray source
is viewed at high inclination through a moderate optical depth ADC.
They predict both a deep Fe-K edge and a prominant K-$\alpha$
emission line - again neither of which are seen. Thus the \src\ spectrum
cannot be easily understood in terms of current models of X-ray production
and reprocessing in 
ADCs and we await future high quality measurements to shed more light on 
this complex spectrum.

\acknowledgements 
The BeppoSAX satellite is a joint Italian and Dutch programme. 
We thank the referee, Martin Still, for helpful comments 
and Ken Ebisawa, Koji Mukai, and the staff of the BeppoSAX Science 
Data Center for assistance. 
This research has made use of data obtained through the High Energy 
Astrophysics Science Archive Research Center Online Service, provided 
by the NASA/Goddard Space Flight Center.

{}

\begin{thebibliography}{}

\bibitem[1989]{a:89}
Anders E., Grevesse N., 1989, Geochimica et Cosmochimica Acta 53, 197 

\bibitem[1997]{b:97}
Boella G., Chiappetti L., Conti G., et al., 1997, A\&AS 122, 327

\bibitem[1997]{f:97}
Frontera F., Costa E., Dal Fiume D., et al., 1997, A\&AS 122, 371

\bibitem[2000]{g:99}
Guainazzi M., Parmar A.N., Oosterbroek T., 2000, Ap. Lett \& Comm. submitted

\bibitem[1989]{h:89}
Hellier C., Mason K.O., 1989, MNRAS 239, 715

\bibitem[1992]{h:92}
Hellier C., Mason K.O., Williams O.R., 1992, MNRAS 258, 457

\bibitem[1994]{h:94}
Hellier C., Smale A.P., 1994. In: Holt S.S., Day C.S.R. (eds.) 
Evolution of X-ray Binaries, AIP Conf. Ser 308, p.~535

\bibitem[1995]{h:95} 
Hua X.-M., Titarchuk L., 1995, ApJ 449, 188

\bibitem[1996]{k:96}
Kallman T.R., Leidahl D., Osterheld A., Goldstein W., Kahn S., 1996,
ApJ 465, 994

\bibitem[1994]{k:94}
Ko Y.-K., Kallman T.R., 1994, ApJ 431, 273

\bibitem[1997]{m:97}
Manzo G., Guarrusso S., Santangelo A., et al., 1997, A\&AS 122, 341

\bibitem[1982]{mc:82}
Mason K.O., C\'ordova F.A., 1982, ApJ 262, 253

\bibitem[1982]{m:82}
Mason K.O., Murdin P.G., Tuohy I.R., Seitzer P., Branduardi-Raymont
G., 1982, MNRAS 200, 793

\bibitem[1983]{m:83}
Morisson D., McCammon D., 1983, ApJ 270, 119

\bibitem[1997]{p:97} 
Parmar A.N., Martin D.D.E., Bavdaz M., et al., 1997, A\&AS 122, 309

\bibitem[1999]{p:99}
Parmar A.N., Oosterbroek T., Guainazzi M., et al., 1999, A\&A 351, 225 

\bibitem[1995]{ps:95}
Predehl P., Schmitt J.H.M.M., 1995, A\&A 293, 889

\bibitem[1995]{p:95}
Puchnarewicz E.M., Mason K.O., C\'ordova F.A., 1995, Adv. Space Res. 16, 3,
65

\bibitem[1994]{t:94}
Tanaka Y., Inoue H., Holt S.S., 1994, PASP 46, L37

\bibitem[1994]{ti:94}
Titarchuk L., 1994, ApJ 434 570

\bibitem[1995]{t:95} 
Titarchuk L., Lyubarskij Y., 1995, ApJ 450, 876

\bibitem[1995]{v:95}
Van Paradijs J.,  McClintock J.E., 1995. In: Lewin W.H.G., van Paradijs J.,  
van den Heuvel E.P.J. (eds.) X-ray Binaries. Cambridge University 
Press, Cambridge, p.~58

\bibitem[1993]{v:93}
Vrtilek S.D., Soker N., Raymond J.C., 1993, ApJ 404, 696

\bibitem[1982]{w:82}
White N.E., Holt S.S., 1982, ApJ 257, 318

\bibitem[1981]{w:81}
White N.E., Becker R.H., Boldt E.A., et al., 1981, ApJ 247, 994 

\bibitem[1988]{w:88}
White N.E., Stella L., Parmar A.N., 1988, ApJ 324, 363

\bibitem[1997]{w:97}
White N.E., Kallman T.R., Angelini L., 1997. In: 
Makino F., Mitsuda K. (eds.)
X-Ray Imaging and Spectroscopy of Cosmic Hot Plasmas,
Waseda University, Tokyo, p.~411
\end{thebibliography}
\end{document}